# Part I: Theoretical Predictions of Preferential Oxidation in Refractory High Entropy Materials

Lavina Backman[a,†], Joshua Gild[b,i], Jian Luo[b,ii], Elizabeth J. Opila[a,iii]

[a]Department of Materials Science and Engineering, University of Virginia, Charlottesville VA 22904

[b]Materials Science and Engineering Program, University of California-San Diego, La Jolla, CA

[†]Corresponding author. Postal address: 395 McCormick Road, P.O. Box 400745, Charlottesville VA 22904. E-mail address: lb2ty@virginia.edu. Phone number: +1 (434) 982 5645.

Abstract

High entropy materials, which include high entropy alloys, carbides, and borides, are a topic of substantial research interest due to the possibility of a large number of new material compositions that could fill gaps in application needs. There is a current need for materials exhibiting high temperature stability, particularly oxidation resistance. A systematic understanding of the oxidation behavior in high entropy materials is therefore required. Prior work notes large differences in the thermodynamic favorability between oxides formed upon oxidation of high entropy materials. This work uses both analytical and computational thermodynamic approaches to investigate and quantify the effects of this large variation and the resulting potential for preferential component oxidation in refractory high entropy materials including group IV-, V- and VI-element based alloys and ceramics. Thermodynamic calculations show that a large tendency towards preferential oxidation is expected in these materials, even for elements whose oxides exhibit a small difference in thermodynamic favorability.  The effect is reduced in carbides, compared to their alloy counterparts. Further, preferential oxidation in high entropy refractory materials could result in possible destabilization of the solid solution or formation of other, competing phases, with corresponding changes in bulk material properties.

[i]jgild@ucsd.edu; [ii]jluo@ucsd.edu; [iii]opila@virginia.edu

1.0 Background and Introduction

High entropy metals and ceramics, which are solid solutions consisting of four or more principal components are a new paradigm in materials design [1–6], and are of interest due to the large number of compositions that can be formulated and explored. Such an expansion of possible compositions can potentially fill challenging application needs, such as materials with improved mechanical and chemical stability at high temperatures. These multiple principal component materials have high configurational entropy, which is hypothesized to stabilize the solid solution phase or provide new combinations of phases. Prior research efforts in high entropy alloys (HEAs), summarized in references [2,7,8], noted their improved mechanical properties. Other hypothesized advantages to multi-principal component, high entropy materials are sluggish diffusion and cocktail effects, a term used to describe the synergistic effects of combining multiple elements in large concentrations [9]. Refractory metal alloys and/or related carbides and borides are the focus of this and a companion experimental paper. Some researchers have explored the high temperature oxidation resistance of refractory HEAs [10,11] and carbides [12], and reported oxide phases formed and oxidation kinetics. However, a more rigorous understanding of the thermodynamic driving forces behind the oxidation of high entropy materials is needed. Further, the effects of oxidation on the stability of the underlying material, the configurational entropy stabilizing the phases and their attendant properties, still need to be further understood to aid in the design of oxidation-resistant high entropy materials.

Assuming a generic oxidation reaction for species M, M + $O_2$ (g) = $MO_2$, the driving force for oxidation is given by the familiar free energy equation,

$$\Delta G^o_{rxn} = \Delta H^o_{rxn} - T\Delta S^o_{rxn} \qquad (1)$$

where the free energy of reaction is given by the difference between the free energies of formation of the products and reactants, as shown in Equation (2).

$$\Delta G^o_{rxn} = \Sigma \Delta G^o_{f,prod} - \Sigma \Delta G^o_{f,reac} \qquad (2)$$

Entropy can be further described as in Equation (3).

$$S = S_{vib} + S_{config} \qquad (3)$$

High entropy materials are characterized by additional configurational entropy due to mixing, which can be calculated using Equation (4):

$$\Delta S_{config} = -R \sum_{i=1}^{N} x_i \log x_i \qquad (4)$$

$R$ refers to the ideal gas constant, $N$ is the number of species, and $x_i$ is the mole fraction of each species. The maximum configurational entropy occurs when all species in the alloy occur at equimolar fractions and is given by Equation (5).

$$\Delta S_{config,max} = R \log N \qquad (5)$$

The determination of the products that form during an oxidation reaction is a necessary step in the design of a multi-component, oxidation resistant, ultra-high temperature system, as discussed in depth in a previous publication [13]. When a multi-component material is exposed to a sufficiently oxidizing environment, it is expected that all species in contact with the environment will initially oxidize. With this assumption and considering Equations (1), (2) and (5), along with the reasonable assumption that a high entropy alloy does not form a high entropy (solid solution) oxide scale and the substrate has additional configurational entropy from mixing, one could write Equation (6) and simplify to Equation (7).

$$\Delta G^o_{rxn,HE} = (\Delta H^o_{f,prod} - \Delta H^o_{f,reac}) - T[\Delta S^o_{f,prod} - (\Delta S^o_{f,reac} + \Delta S_{config,reac})] \qquad (6)$$

$$\Delta G^o_{rxn,HE} = \Delta G^o_{rxn} + T\Delta S_{config,reac} \qquad (7)$$

Where $\Delta S_{config,reac}$ is the additional configurational entropy due to the formation of the high entropy solid solution in the substrate. Any changes in vibrational entropy due to the mixing are assumed to be negligible, as in prior analysis [7]. In Equations (6) and (7), $\Delta G_{rxn,HE}^o$ is the driving force for oxidation from a high entropy material, assuming a high entropy oxide does not form. Equation (7) considers the effect of configurational entropy on the driving force for oxidation, and indicates that configurational entropy has the potential to reduce the driving force for oxidation, i.e., $\Delta G_{rxn}^o$ becomes less negative, however the magnitude of this reduction is minimal, as discussed below. The magnitude of this reduction is even less if configurational entropy due to mixing in the oxide is considered.

Figure 1 shows substantial differences in relative thermodynamic favorabilities of the oxides of the group IV, V and VI transition metals. The enthalpy for formation of $HfO_2$ (most favorable) at 2073K is −1115.8kJ/mol and $(WO_3)_3(g)$ (least favorable) is −438.1kJ/mol. The reduction in driving force due to the contribution of the configurational entropy is at least an order of magnitude less than the relative differences in the enthalpy of oxidation. For a five-component equimolar solution, $S_{config}$=1.61R. At 2073K, a relevant temperature for refractory materials, this results in a $T\Delta S$ of +27.7kJ, and the $\Delta G_{rxn,HE}^o$ [Equation (7)] becoming more positive by this amount. It is therefore not expected that the additional configurational entropy will provide a significant advantage in reducing the driving force for oxidation. It is further hypothesized that given the number of components and the large differences in thermodynamic favorability of the oxides of interest, that a large tendency towards preferential oxidation will be exhibited in refractory high entropy materials. Both a first-principle surface oxidation study [14] and an experimental study on bulk material [11] have observed preferential oxidation consistent with the thermodynamic favorability indicated in Figure 1. This demonstrates the need for a deeper understanding of the extent to which this phenomenon can be expected to occur in high entropy materials. Preferential oxidation also has implications for the overall stability of the underlying high entropy material, leading to destabilization of the solid solution phase or formation of undesirable phases, due to the reduction in configurational entropy in the initial material as some components are preferentially depleted.

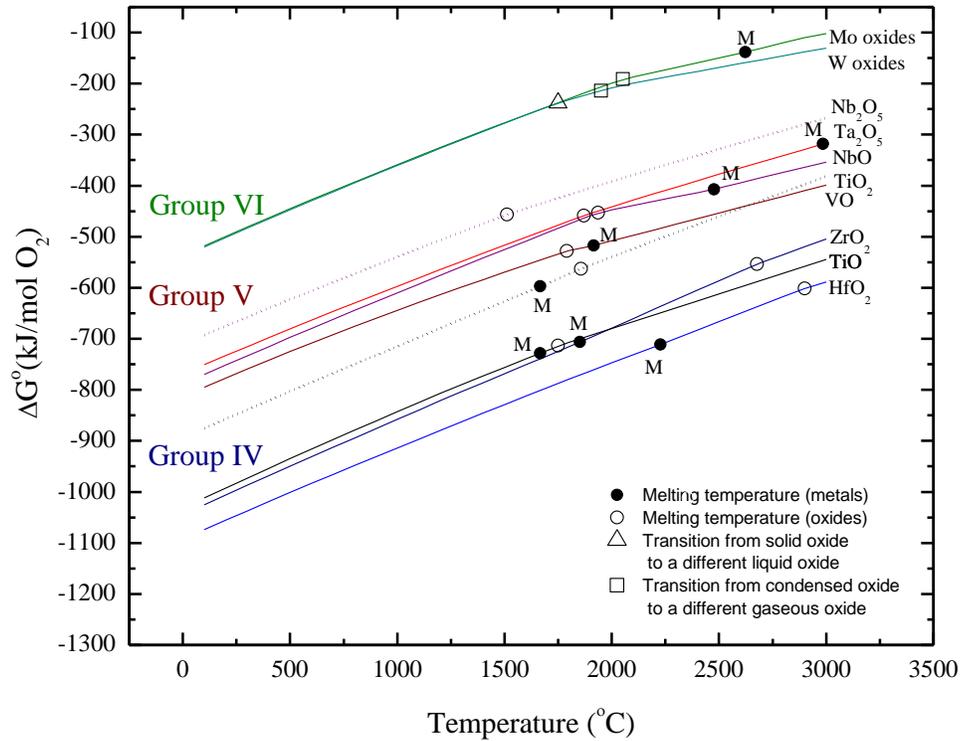

**Figure 1**: Ellingham diagram showing the reaction free energies per mol $O_2$ of the group IV, V and VI refractory metal oxides. Filled circles indicate the melting temperatures of the metals; open circles indicate melting temperatures of the oxides. Adapted from reference [13], with $Nb_2O_5$ and $TiO_2$ added in dashed lines.

This work used a thermodynamic approach to provide a quantitative analysis of preferential oxidation in multi-principal component refractory materials as a function of elemental composition. These results were compared to experimental results in a companion publication. The calculations were idealized, first to simplify the complex compositional and phase assemblages, and second, because thermodynamic data were not available for these multicomponent systems and complex oxides. The overall objective of this work was to predict the oxidation products of multi-principal component materials and thereby provide insight into the mechanisms that govern their oxidation behavior.

## 2.0 Methodology

### 2.1 Analytical Approach Using Equilibrium Constants

The equilibrium composition of the oxide scale formed from multi-component base materials was sought. The equilibrium oxide formed will not have the same composition (on a cation basis) as the underlying material due to varying constituent thermodynamic favorabilities. Consider a binary Hf-Zr alloy. One might expect this alloy to oxidize according to the reaction:

$$(Hf_{0.5}Zr_{0.5}) + O_2\ (g) \leftrightarrow (Hf_{0.5}Zr_{0.5})O_2 \tag{8}$$

Equation (8) suggests that a 50-50 (at%) Hf-Zr alloy will yield a 50-50 (at%) solution of $HfO_2$ and $ZrO_2$. However, $HfO_2$ is more thermodynamically favored compared to $ZrO_2$ ($\Delta G^o_{f,HfO_2}$=−779.4 kJ/mol and $\Delta G^o_{f,ZrO_2}$=−715.6 kJ/mol at 2073K), and the remaining Hf could reduce any $ZrO_2$ that forms. This difference in thermodynamic favorability will be reflected in the resulting oxide composition. When an Hf-Zr alloy oxidizes, it will be in equilibrium with a solid solution of $HfO_2$ and $ZrO_2$, with a corresponding equilibrium $pO_2$. The relevant reactions for each alloy constituent are:

$$Hf + O_2(g) \leftrightarrow HfO_2 \tag{9}$$

$$Zr + O_2(g) \leftrightarrow ZrO_2 \tag{10}$$

Following Gaskell's approach [15], the pertinent reaction describing the equilibrium between the base metal constituents and their oxides can be written by combining Equations (9) and (10) as

$$Hf + ZrO_2 \leftrightarrow HfO_2 + Zr \tag{11}$$

Assuming ideal solid solution thermodynamics, where each activity coefficient, $\gamma$, is equal to unity, the equilibrium constant for this equation yields the following relationship for the relative concentrations in mole fraction ($X_i$) of the components

$$K = \frac{[X_{HfO_2}][X_{Zr}]}{[X_{ZrO_2}][X_{Hf}]} \tag{12}$$

For a fixed temperature, *K* is constant; therefore, for a given alloy composition, e.g. a 50-50 alloy, the composition of the oxide will be fixed as well.

$$K * \frac{[X_{Hf}]}{[X_{Zr}]} = \frac{1 - [X_{ZrO_2}]}{[X_{ZrO_2}]} \quad (13)$$

Using Equation (13), the relative mole fractions of oxides in equilibrium with a range of Hf-Zr alloy compositions can be plotted at a given temperature in a *composition balance diagram*. Using tabulated data, e.g. FactSage databases [16], to determine *K*, the analysis shown in Equations (11)-(13) was also conducted for Zr-Ti, Ti-Ta and Ta-Mo, in equilibrium with the predominantly observed, condensed binary oxides in the respective systems. These elements were chosen to represent the typical elements that make up refractory high entropy materials. In the case of the Ta-Mo alloy, $MoO_2$ was chosen as the representative oxide since it is the thermodynamically stable condensed oxide in the Mo-O system [13], and has been observed to exist in the sublayers of thermally grown oxides or other conditions with reduced oxygen partial pressures [17,18]. At higher temperatures and oxygen partial pressures, the highly volatile, less thermodynamically favored $MoO_3$ (l) is formed [19].

These analyses were then extended to the more complex refractory carbide systems. Equations (14) and (15) demonstrate the governing reactions and the resulting equilibrium constant, using the example of a TiC-TaC solid solution. Note the formation of the additional product, CO (g).

$$TiC + \frac{3}{7}Ta_2O_5 \leftrightarrow TiO_2 + \frac{6}{7}TaC + \frac{1}{7}CO \;(g) \quad (14)$$

$$K_c = \frac{[X_{TiO_2}][X_{TaC}]^{\frac{6}{7}} pCO^{\frac{1}{7}}}{[X_{TiC}][X_{Ta_2O_5}]^{\frac{3}{7}}} \quad (15)$$

Oxygen solubility in the metallic systems was not considered, which can be as high as 30%, as in the Ti-O system [20]. Similarly, oxygen solubility in the rocksalt carbide lattice [21] was also neglected.

Oxygen solubility in high entropy alloy and ceramic systems and the thermodynamic effects of this dissolution have not been studied [22], and are therefore unknown. Experimental studies were undertaken to gauge the applicability of these tools to real systems, and the results reported in a companion paper. It is expected that this thermodynamic assessment method [Equations (11)-(15)] will assist with oxide phase identification and prediction of preferential oxidation, which are essential for understanding the oxidation mechanisms of refractory high entropy materials, and thereby aid design efforts for oxidation-resistant materials.

*2.2 Free Energy Minimization Calculations*

The analytical method outlined by Gaskell is a useful way to consider the binary ideal solution case, and provided a basis for the development of hypotheses with respect to preferential oxidation in real, multicomponent cases. These hypotheses were further refined by leveraging the computational capabilities of FactSage, a free energy minimization software package [16], and the accompanying databases. The FactSage Equilibrium module allows for the definition of solid solution reactant and product phases, both ideal and real, provided the pertinent databases are available. In both the real and ideal cases, the appropriate phase diagrams, mostly pseudo-binary, were consulted to identify the most likely phases and/or components of the solution at the temperature under consideration. Current research is addressing the gaps for ternary oxides [23,24], but further information on solid solubility is needed.

Similar to the analytical approach in the prior section, an equimolar solid solution substrate (alloy or carbide) was defined and considered to be in equilibrium with an equimolar oxide solid solution (on a cation basis). The assumption of ideal solution thermodynamics, unless otherwise stated, was made to approximate solid solutions formed from phases of differing crystal structures (as input), due to the data available in the Fact Pure Substance database. The Gibbs free energy of the assemblage of substrate and oxide constituents was then minimized, with a resulting output of equilibrium phases. In cases where sub-oxides were predicted to form, the calculation was repeated using suboxides as additional input in the oxide solid solution phase. The equilibrium $pO_2$ between the resultant oxide and substrate is also calculated. The case of the Hf-Zr binary alloy oxidizing to $HfO_2$-$ZrO_2$ in an ideal solid solution phase was compared to results from free energy minimization calculations for the same system in FactSage to test this methodology. Once tested, the approach was extended to study ternary (Hf-Zr-Ti) and quinary (Hf-Zr-Ti-Ta-Nb) alloy compositions. The results from these calculations reflect the local equilibrium near the oxide-substrate interface and are therefore descriptive of the oxidation affected zone.

The ideal solution assumptions described above allowed for a first approximation in the understanding of the tendency towards preferential oxidation in high entropy materials, which will be shown in the companion paper to have good agreement with experimental results. This simplified approach relied on thermodynamic data available in the FactSage database for pure substances, and assumed that all oxide compounds will go into solution without applying an enthalpy correction for a change in crystal structures. The activity of the oxide solution phase was constrained to be unity. This approximation imposed the coexistence of, for instance, a rutile Ti oxide constituent in solution with the tetragonal Hf and Zr oxides; without this constraint, the free energy minimization calculations would have resulted in the most stable product phase being $HfO_2$, i.e., $HfO_2$ as having an activity of one, and the other oxide phases as having been reduced by Hf. This approach was reasonable as the enthalpy correction would have been negligible compared to the free energy of formation of the oxides, or the relative thermodynamic favorabilities (relative values of $\Delta G_{rxn}^o$) of the oxides of interest [25–27].

Additionally, in non-ideal thermodynamics, the activity of a component in a solution varies from the mole fraction according to the activity coefficient, $\gamma$. While this non-ideality would affect the reactant solid solution as well as the overall reaction free energy, it is hypothesized these differences would not be so large compared to the relative differences in the ΔG° for oxidation between oxides from different groups (Figure 1). These large differences were expected to persist when comparing the oxidation free energies using thermodynamic data for real solutions. Based on these considerations, the use of ideal solution thermodynamics is sufficient to understand the preferential oxidation phenomenon in HE materials. To probe this, the case of an ideal Zr-Ti alloy to form real oxidation products, rather than ideal products, was considered. The Zr-Ti system was chosen due to the availability of thermodynamic data for their oxides. While Zr and Ti metals are soluble in each other, the binary phase diagram for their oxides exhibits more complex phase behavior [28]. Their oxides also exhibit a large difference in their relative thermodynamic favorabilities. The calculation was performed at the Zr-Ti equimolar alloy composition at 1773K, assuming the alloy to be an ideal solution, where the temperature was chosen below the melting temperature of Ti. The FTOxid database, which contains thermodynamic data for real oxide solutions such as $(Zr,Ti)O_2$ tetragonal oxide solution phase and a rutile oxide phase, was used to calculate the oxide composition in the non-ideal case.

The focus of this work was on equimolar multi-principal component materials. Temperatures used in the following calculations were chosen to be as high as possible while maintaining all constituents in the solid state. Solid states were chosen over liquid so that temperatures remained within the realm of experimental verification described in Part II. Partial pressures (oxygen activity) reported in this study were those that were predicted to exist between the external oxide scale and substrate at equilibrium.

## 3.0 Results

*3.1 Composition Balance Diagrams: Ideal Binary Metal Alloys and Ideal Oxide Solutions*

The analytical approach described in section 2.1 enables the generation of composition balance diagrams, which are shown in Figures 2-6. Figures 2-4 show the composition balance diagrams for Hf-Zr, Ti-Ta and Ta-Mo assuming ideal solution behavior in both binary metal alloys and their oxides. Results for the formation of real oxide solutions ($\gamma \neq 1$) are presented in section 3.4. The composition balance diagram can be read as follows. Take for example Figure 2, the top x-axis, shown in mole fraction, ranges from pure $HfO_2$ on the left side to pure $ZrO_2$ on the right. The bottom x-axis, also in mole fraction, ranges from pure Hf on the left to pure Zr on the right. The y-axis height is set arbitrarily to display the slopes of the tie lines joining the metal alloy composition with the corresponding equilibrium oxide composition, assuming that both alloy and the oxides are in an ideal solid solution respectively. The tie lines are also oxygen isobars, representing the equilibrium partial pressure of oxygen (based on a reference state of 1 atm) that corresponds to each specific alloy-oxide activity ratio. The lines joining the pure Hf to pure $HfO_2$, and that joining pure Zr to pure $ZrO_2$, correspond to the equilibrium $pO_2$ for the respective oxidation reactions, given by Equations (9) and (10).

Figure 2 shows the case of an alloy consisting of two elements from the same group in the periodic table, Hf and Zr, whose oxides have similar thermodynamic favorability (Figure 1). The diagram shows that more than 97% zirconium in the alloy is needed to form ~50% $ZrO_2$. Taking the equimolar case, a 50-50 mol% Hf-Zr alloy will yield at equilibrium an oxide scale with 2.4% $ZrO_2$ (Figure 2). If one follows the red dotted line on Figure 2 upward, each intersection with a black tie-line represents a higher partial pressure of oxygen. When the alloy is in its initial state, the composition is 50 mol% Zr. At this composition, the equilibrium partial pressure (oxygen activity) between the alloy and the oxide scale is $4\times10^{-20}$ (with respect to 1 atm), and the oxide composition is 2.4 mol% $ZrO_2$. Correspondingly, as one moves upwards along the red dotted line, an oxide increasingly enriched in the less stable oxide, in this case $ZrO_2$, is encountered. As Hf preferentially oxidizes, and is depleted from the base alloy, the alloy composition will traverse to the right on the lower x-axis in the depletion zone. This results in traversing to the right on the top x-axis (oxide composition), i.e., assuming an inward growing oxide, the Hf-depleted regions become richer in Zr, forming more $ZrO_2$, with a concomitant increase in the equilibrium partial pressure of oxygen.

Figures 3 and 4 represent the oxidation of alloys in which the constituents are members of adjacent groups in the periodic table with a much larger difference in the free energies required to form their respective oxides. In both these cases, the oxide composition is comprised entirely or almost entirely by the element with the most thermodynamically favored oxide.

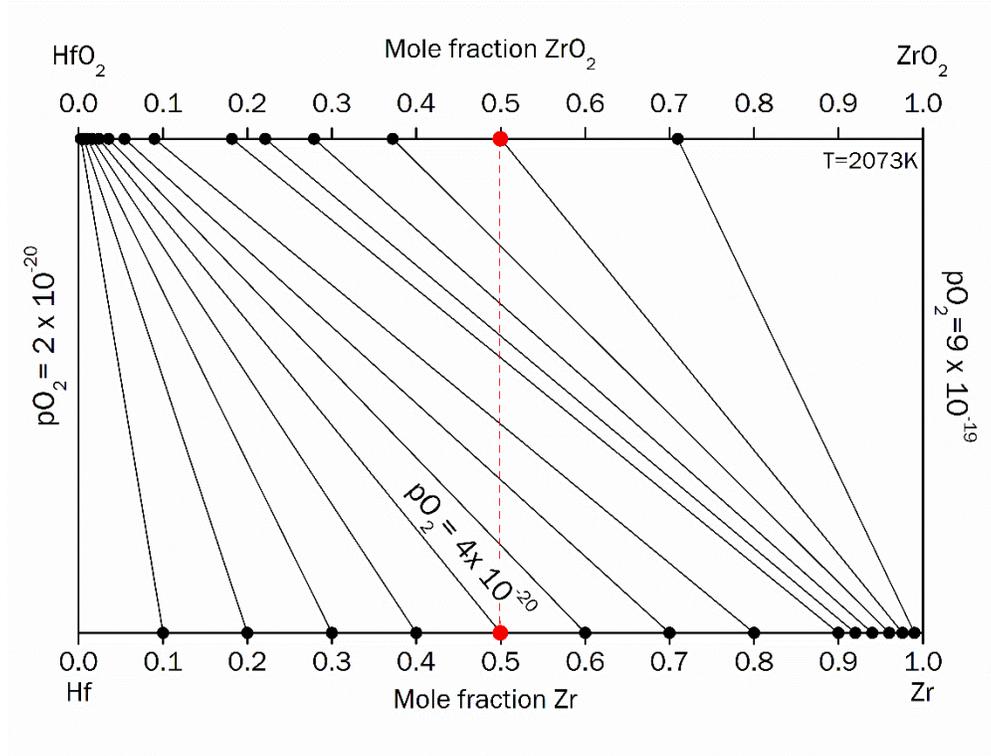

**Figure 2**: Composition balance diagram showing the equilibrium concentrations of the oxides (tetragonal phase) in a binary solution when formed from a binary metal alloy, Hf-Zr (group IV). Results are shown in increments of 10 mol % with respect to alloy composition at 2073K. The red dotted line is a *reference* line, which goes from the 50 mol% Zr composition in the alloy to the 50 mol % $ZrO_2$ in the oxide.

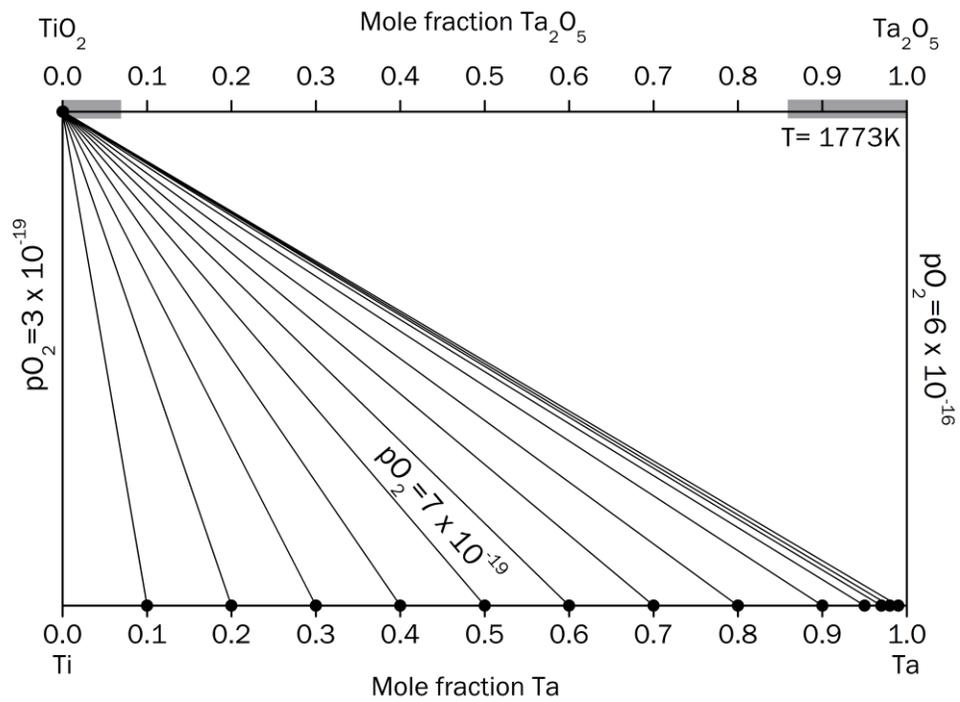

**Figure 3**: Composition balance diagram showing the equilibrium concentrations of the oxides in a binary solution when formed from a binary metal alloy, Ti-Ta (groups IV and V), of varying compositions at 1773K. Note that even at 99% tantalum in the alloy, the oxide is expected to be composed of only 0.08% $Ta_2O_5$ at equilibrium. $TiO_2$ is assumed to be in the rutile phase. Grey shaded areas on the oxide axis indicate regions of solid solubility in the real system according to reference [29].

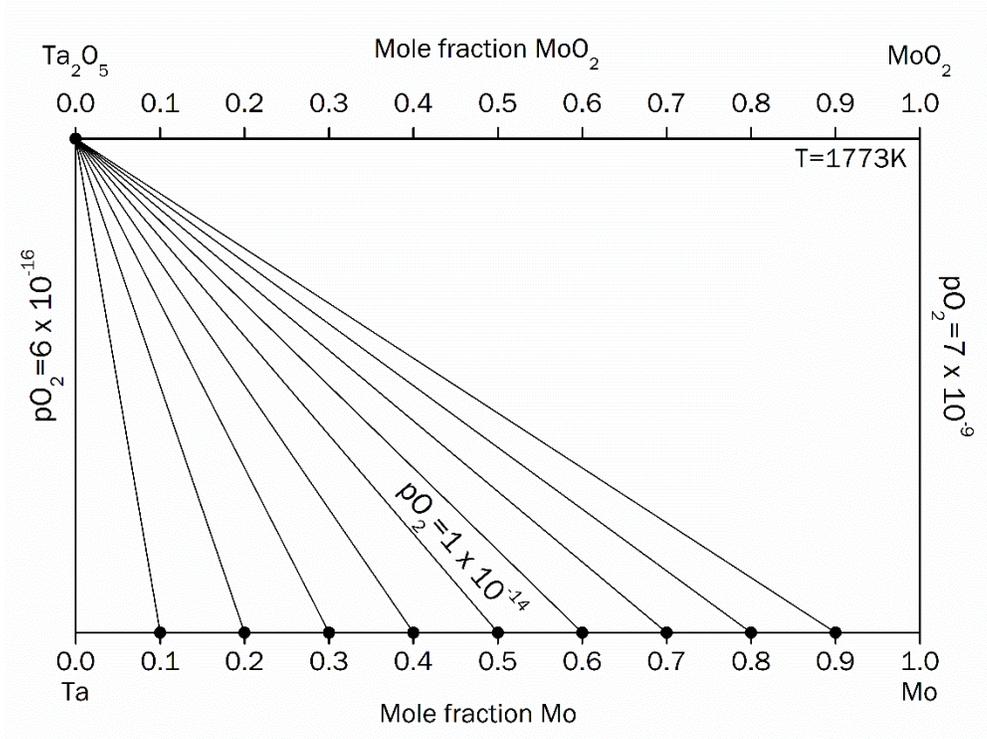

**Figure 4**: Composition balance diagram showing the equilibrium concentrations of the oxides in a binary solution when formed from a binary metal alloy, Ta-Mo (groups V and VI), of varying compositions at 1773K. Note that the larger difference in thermodynamic favorability in the oxides when comparing different groups (group V vs VI) results in a large tendency towards preferential oxidation to form $Ta_2O_5$, the more thermodynamically favored oxide.

*3.2 Composition Balance Diagrams: Ideal Transition Metal Carbide and Ideal Oxide Solution*

A similar approach was followed to calculate the composition balance diagram for the refractory carbides, where the additional product, CO (g), must be considered. In the case of HfC-ZrC, which would oxidize to $HfO_2$ and $ZrO_2$ respectively, the condensed phases have the same oxygen stoichiometry and the pCO term cancels out. Results are shown in Figure 5. Note that for the HfC-ZrC oxidizing to $HfO_2$-$ZrO_2$ (Figure 5), the oxide solution is more reflective of substrate composition (on a cation basis), compared to the metallic alloy case (Figure 2). Note as well that the equilibrium pO$_2$s are the same order of magnitude and are both significantly higher than those of the end members in the alloy case (Figure 2). For TiC-TaC:TiO$_2$-Ta$_2$O$_5$ system, the pCO term remains (Equation (15)). To simplify the analysis, the value of pCO was fixed at 1. The resulting diagram is shown in Figure 6. For TiC-TaC oxidizing to TiO$_2$-Ta$_2$O$_5$ (Figure 6), the tendency towards preferential formation of TiO$_2$ is also decreased relative to the metallic alloy case (Figure 3). However, TiO$_2$ is still strongly favored over Ta$_2$O$_5$.

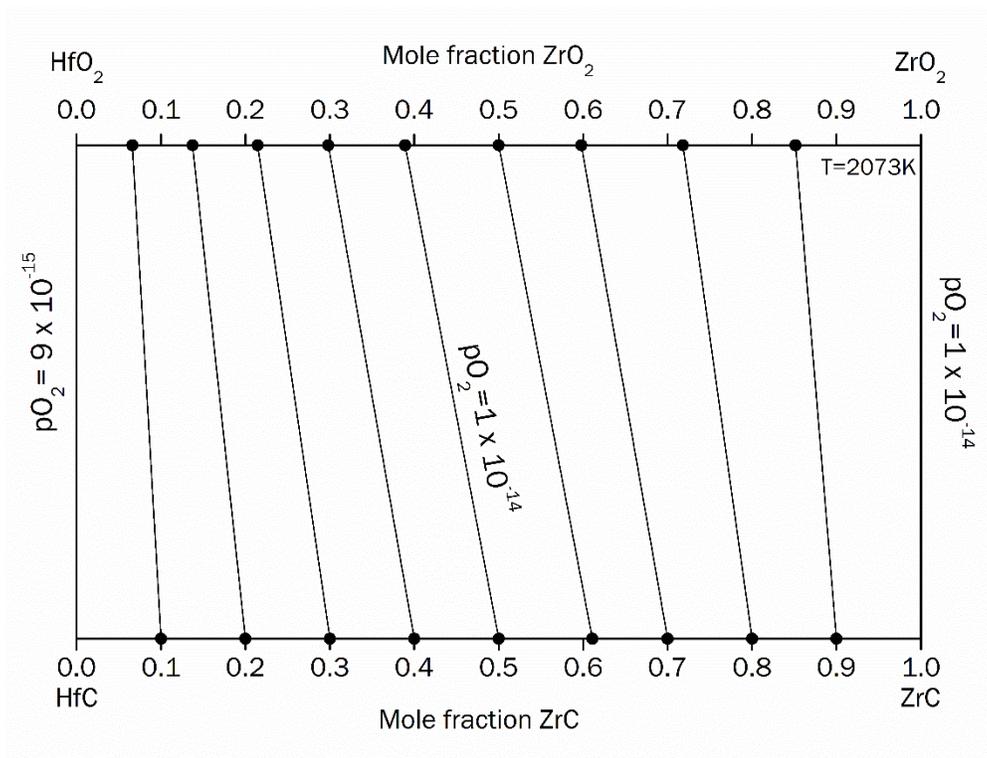

**Figure 5**: Composition balance diagram showing the equilibrium concentrations of the oxides in a binary solution when formed from a binary carbide, HfC-ZrC, of varying compositions at 2073K.

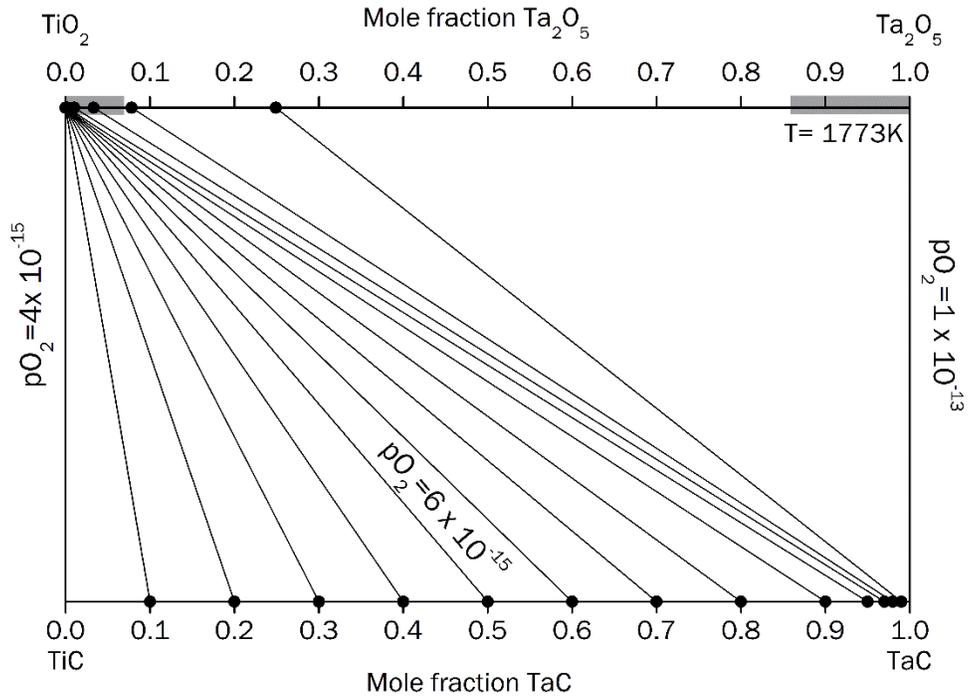

**Figure 6**: Composition balance diagram showing the equilibrium concentrations of the oxides in a binary solution when formed from a binary carbide, TiC-TaC, of varying compositions. Note that the tendency towards preferential formation of $TiO_2$ is decreased relative to the metallic alloy case (Figure 3). pCO was assumed to be 1. $TiO_2$ is assumed to be in the rutile phase. Grey shaded areas on the oxide axis indicate regions of solid solubility in the real system according to reference [29].

*3.3 Free Energy Minimization: Ideal Ternary and Quinary Alloy and Ideal Oxide Solutions*

The approach using the FactSage free energy minimization software were first confirmed by comparison to the analytical results for the Hf-Zr system (Figure 2). Thus validated, this approach was then used to generate predictions for systems with more than two components. The equimolar cases for a ternary system, Hf-Zr-Ti, and a quinary system, Hf-Zr-Ti-Ta-Nb, at 1773K were studied as examples. Both the ternary and quinary alloys under consideration are known to be in the BCC phase at this temperature [30–32]. In the ternary case, the HfZrTi alloy was assumed to be in equilibrium with an ideal solid solution containing $HfO_2$, $ZrO_2$ and $TiO_2$. Similarly, the HfZrTiTaNb alloy was assumed to be in equilibrium with an ideal solid solution containing $HfO_2$, $ZrO_2$, $TiO_2$, $Ta_2O_5$ and $Nb_2O_5$. There are few data on the quinary phase equilibria for the corresponding oxide phases; therefore, the most stable, condensed phase for each oxide according to available data was assumed, regardless of crystal structure.

Tables 1 and 2 summarize the results of these calculations. In these tables, the input refers to the initial compositions of the oxide and alloy solid solutions to be placed in equilibrium with each other. The output refers to the final compositions of each, after free energy is minimized. In the ternary case (Table 1), the final composition of the oxide is primarily of Hf with some Zr and very little Ti; the alloy is depleted in Hf. In the quinary case (Table 2), note the preferential formation of group IV oxides (Hf, Zr, Ti) and corresponding depletion of those constituents in the alloy. Ti suboxides (TiO, $Ti_2O_3$) are also predicted to form.

**Table 1**: Equilibrium calculation showing the final composition of the oxide scale that results from an equimolar HfZrTi ternary alloy at 1773K in equilibrium with an equimolar oxide ideal solid solution containing $HfO_2$, $ZrO_2$ and $TiO_2$, as well as depletion of Hf from the underlying alloy.

| Constituents | Input (mol) | Input (mol%) | Output (mol %) |
|---|---|---|---|
| **Alloy, Ideal Solution** | | | |
| Hf | 1 | 33.3 | 1.03 |
| Zr | 1 | 33.3 | 32.3 |
| Ti | 1 | 33.3 | 66.7 |
| **Oxide, Ideal Solution** | | | |
| $HfO_2$ | 1 | 33.3 | 65.6 |
| $ZrO_2$ | 1 | 33.3 | 34.4 |
| $TiO_2$ | 1 | 33.3 | $4.70 \times 10^{-3}$ |

**Table 2**: Equilibrium calculation showing the final composition of the oxide scale that results from an input of equimolar quinary $Hf_{0.2}Zr_{0.2}Ti_{0.2}Ta_{0.2}Nb_{0.2}$ alloy at 1773K in equilibrium with an equimolar (cation atom basis) oxide ideal solid solution containing $HfO_2$, $ZrO_2$, $TiO_2$, $Ta_2O_5$, $Nb_2O_5$, as well as depletion of Hf, Zr and Ti from the underlying alloy.

| Constituents | Input (mol) | Input (mol%) | Output (mol %) |
|---|---|---|---|
| **Alloy Ideal Solution** | | | |
| Hf | 1 | 20 | $7.87 \times 10^{-8}$ |
| Zr | 1 | 20 | $4.72 \times 10^{-6}$ |
| Ti | 1 | 20 | $8.05 \times 10^{-3}$ |
| Ta | 1 | 20 | 48.4 |
| Nb | 1 | 20 | 51.5 |
| **Oxide Ideal Solution** | | | |
| $HfO_2$ | 1 | 20 | 35.9 |
| $ZrO_2$ | 1 | 20 | 35.9 |
| $TiO_2$ | 1 | 20 | 4.05 |
| Ti sub-oxides (TiO, $Ti_2O_3$) | - | - | 23.2 |
| $TaO_{2.5}$ | 1 | 10 | 1.05 |
| $NbO_{2.5}$ | 1 | 10 | $6.32 \times 10^{-5}$ |

*3.4 Free Energy Minimization: Ideal Alloy and Real Oxide Solution*

Thus far, both the alloy and oxide were assumed to be in ideal solid solutions respectively. Comparison of the predictions from the ideal case for a given system to the real case (using solid solution data from the FTOxid database) offer insight into the suitability of the ideal solution approximation. Table 3 compares the resulting oxide compositions calculated using the ideal solid solution approximation, and real solid solution data, for the oxide solid solution. In both cases, the alloy was assumed to be an ideal solid solution. Using the real oxide solution data, the predicted oxide is tetragonal $ZrO_2$ with small amounts of $TiO_2$, similar to that predicted by the ideal solution approximation.

**Table 3:** Predicted final compositions after starting with an equimolar Zr-Ti ideal alloy solution in equilibrium with an equimolar $ZrO_2$-$TiO_2$ *ideal* solid solution (Case 1) versus an equimolar Zr-Ti ideal alloy solution in equilibrium with an equimolar $ZrO_2$-$TiO_2$ *"real"* solution using data from the FTOxid database (Case 2) at 1773K. Note that in *both* cases, the alloy is assumed to be an ideal solid solution. Note also that partial depletion of Zr from the underlying alloy is predicted in both cases.

|  | **Case 1** | **Case 2** |
| --- | --- | --- |
|  | **Predicted ideal alloy solution composition (mol%)** | **Predicted ideal alloy solution composition (mol%)** |
| **Zr** | 24.6 | 22.2 |
| **Ti** | 75.4 | 77.8 |
|  | **Predicted oxide composition (mol%) assuming an <u>ideal</u> solution in equilibrium with the ideal alloy above** | **Predicted oxide composition (mol%) assuming a <u>"real"</u> solution in equilibrium with the ideal alloy above** |
| **$ZrO_2$** | 99.9 | 99.9 |
| **$TiO_2$** | .020 | .0197 |

## 4.0 Discussion

*4.1 Composition Balance Diagrams: Binary Metal Alloys*

The composition balance diagrams shown in section 3 provided a visual method to illustrate the extent of preferential oxidation expected. They also demonstrate the strong sensitivity of preferential oxidation to the difference in relative thermodynamic favorability between the oxides of interest. This sensitivity is best seen when comparing Figures 2 and 3 where the periodic trend in the thermodynamic favorabilities of the oxides of the elements under consideration is shown. $HfO_2$ and $ZrO_2$, formed from elements in the same periodic table group, have similar free energies of formation, normalized to one mole of $O_2$ ($\Delta G^o_{HfO_2}$= -779.4 kJ/mol and $\Delta G^o_{ZrO_2}$= -715.6 kJ/mol at 2073K), as shown in Figure 1. Even so, this small difference had a large effect on the thermodynamic prediction for preferential oxidation (Figure 2). Given the result for Hf-Zr alloy oxidation which forms two similarly thermodynamically favored oxides, it was not surprising to see that for the oxidation of a Ti-Ta alloy (group IV-V alloy), the tendency towards preferential oxidation of Ti was greater. The difference in the free energy for oxidation per mole of $O_2$ was smaller when comparing between two elements in the same group, Hf-Zr, versus two elements in different groups, Ti-Ta. This is again seen in Figure 4, wherein Ta in Ta-Mo alloy was expected to dominate the composition of resulting oxide scale, even for alloy compositions rich in Mo.

It is to be noted that in the cases shown in Figure 3 and 4, the choice of oxides to represent the end members of the oxide side of the diagram were based on the predominantly observed oxide in each M-O system, where M = Ti, Ta. The existence of suboxides and complex oxides in the Ti-Ta-O system may result in a more complex scale. However, it is unlikely that consideration of these oxides will result in a significantly reduced tendency towards preferential oxidation, given the large difference in thermodynamic favorabilities between their oxides. Park and Butt [33] experimentally studied the oxidation of a Ti-Ta alloy at varying compositions from 0-60 wt% Ta, and found that generally speaking, the oxide scale consisted of a $TiO_2$ scale with increasing Ta content as one moved towards the metal-oxide interface, suggesting $Ta_2O_5$ formed when Ti was depleted from the alloy. This is consistent with the diagram shown in Figure 3 wherein, as more Ti is depleted due to oxidation, the alloy composition becomes Ta-rich, allowing for the formation of Ta-rich oxides.

It is also interesting to compare the composition balance diagram for the metal alloys oxidizing to that of their corresponding carbides oxidizing. Take the case of HfC-ZrC (Figure 5), wherein the oxide composition now more closely reflects that of the base material composition compared to the oxide composition for the corresponding alloy (Figure 2). The spread of the tie lines indicates that preferential oxidation of the HfC component in the carbide solid solution will be much less significant than Hf in the Hf-Zr alloy (50 mol% ZrC yields a 39 mol% $ZrO_2$ oxide compared to 50 mol % Zr yielding 3 mol% $ZrO_2$). This was also the case for TiC-TaC where again a larger spread in the tie lines compared to the metal alloy case was observed. A prior publication [13] discussed the relative differences in free energy for oxidation when considering the alloy, carbide, and boride cases for any two given elements, and the relevant diagram reproduced here (Figure 7). The relative difference in the free energies of the reactions to form oxides from Hf and Zr carbides is almost negligible compared to the metal case, due to the formation of the gaseous oxide, CO (g) and the increase in equilibrium $pO_2$. The difference in the oxide formation energy from $HfB_2$ and $ZrB_2$ fall between these values (Figure 7). Thus, the oxide formation from the metal and the carbide can serve as bounding cases for the purposes of predicting preferential oxidation.

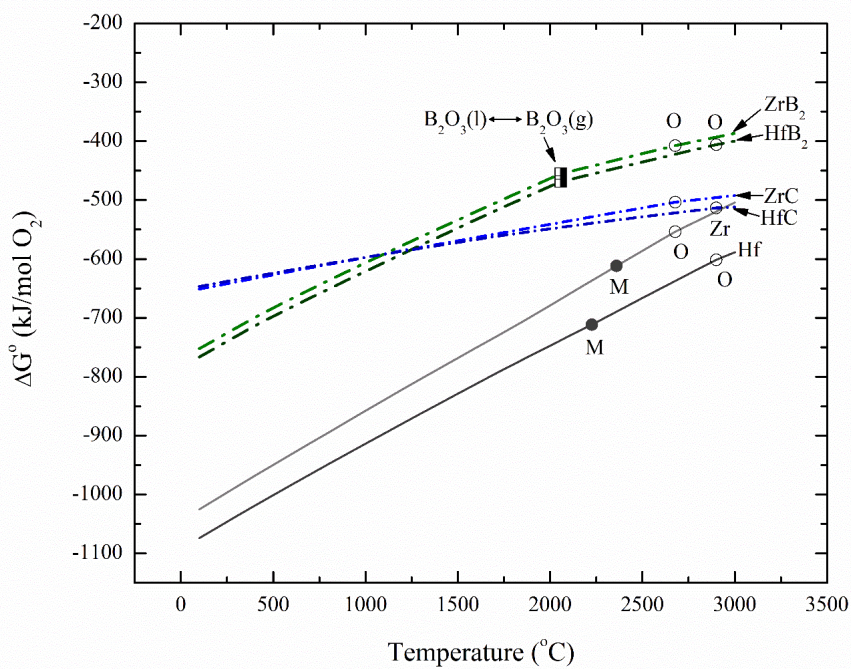

**Figure 7**: Comparison of oxidation reaction free energies to form $HfO_2$ and $ZrO_2$ for pure metal, carbide or boride reactants, respectively. Open circles indicate oxide melting temperatures, and closed circles indicate metal melting temperatures. Adapted from reference [13].

*4.2 Free Energy Minimization*

Calculations for the ternary case using the ideal solid solution approximation showed that the oxidation of the equimolar group IV HfZrTi alloy would result in a $HfO_2$-$ZrO_2$ rich scale. Very small amounts of $TiO_2$ were predicted in the final state, well within the solubility limits in both $HfO_2$ and $ZrO_2$ [28,34]. If the trend exhibited in the ternary case were extrapolated to the high entropy case $Hf_{0.2}Zr_{0.2}Ti_{0.2}Ta_{0.2}Nb_{0.2}$, insignificant amounts of Ta and Nb oxides would be expected. Indeed, preferential oxidation of the group IV elements is predicted, and the oxide scale contains 1 mol% or less of both $Ta_2O_5$ and $Nb_2O_5$ (Table 3). The equilibrium $pO_2$ at the oxide/alloy interface due to the assemblage of the different oxides that were expected to form in the initial stages of oxidation ($HfO_2$,

TiO$_2$, ZrO$_2$, Ta$_2$O$_5$, Nb$_2$O$_5$) was higher than that due to only the most stable oxides (e.g. HfO$_2$, ZrO$_2$), and influenced the final oxide composition determined via the free energy minimization approach.

*4.3 Ideal Solution vs Real Solution Thermodynamics*

The tendency towards preferential oxidation was highly sensitive to the relative difference in the standard Gibbs free energy of formation for the oxides. Figure 1 shows a sizable difference in the relative thermodynamic favorabilities of ZrO$_2$ and TiO$_2$, similar to the difference between Ta$_2$O$_5$ and TiO$_2$ (Figures 1 and 3). This suggests that for a ZrTi alloy, preferential oxidation of Zr is expected. Calculations using both the ideal solid solution approximation and real solid solution data indeed showed that the oxide composition is dominated by ZrO$_2$ (Table 3). Gaps in current knowledge and available data drove the need to assume ideal thermodynamic behavior for most of the calculations presented here. The result in Table 3 confirmed that the considerations regarding the ideal solution approximation outlined in section 2.2 were reasonable, and that this approach can sufficiently predict the extent of preferential oxidation in refractory materials.

*4.4 Effect on Substrate at the Oxidation Interface*

An estimated reduction in configurational entropy due to the reduced number of components in the solid solution alloy or compound were calculated using Equations 4 and 5. For the case shown in Table 2 calculated at 1773K, where the alloy effectively goes from N=5 to N=2 due to preferential oxidation, this resulted in a change in S$_{config}$ from 13.4 J/mol·K to 5.8 J/mol·K (using Equation (4)), which is an approximately 60% reduction in configurational entropy. If configurational entropy is required to stabilize the alloy solid solution phase [35], then such a reduction may result in the destabilization of this phase at the substrate-oxide interface or the oxidation affected zone, and the formation of other, competing phases. This would be accompanied by a corresponding change in material properties. For example, the preferential oxidation of group IV elements in the high entropy carbide may result in the formation of secondary phases such as $Ta_2C$ and $Nb_2C$, which have different crystal structures compared to the original rocksalt. On the other hand, if the remaining components form a thermodynamically stable solid solution, then the effects may include a reduction in melting temperature or in the originally enhanced mechanical properties. The work presented here and previously [13] indicates that such preferential oxidation of refractory alloys and compounds is closely correlated with groups in the periodic table; if the remaining elements all belong to the same periodic group, they may remain in solid solution.

5.0 Conclusions

Periodic trends in the free energies of oxidation of the group IV, V and VI refractory elements were expected to drive preferential oxidation in refractory high entropy materials. An analytical approach for binary systems was extended through the development of a computational approach using free energy minimization. This methodology was used to quantify the extent of preferential oxidation in binary, ternary and quinary alloy and carbide systems. The results indicated that the tendency for preferential oxidation is highly sensitive to the difference in relative thermodynamic stabilities of the constituent oxides. The differences in relative thermodynamic stabilities of oxides from the same periodic group, though small, were enough to result in a significant tendency towards preferential oxidation; this effect was dramatically more pronounced in systems with elements from different periodic groups.

These findings can help elucidate oxidation mechanisms in refractory high entropy materials (alloys, carbides, borides), and provide guiding principles for their design. Preferential depletion of the starting materials' constituents according to the relative thermodynamic favorability of their respective oxides will reduce the effect of configurational entropy in stabilizing the material in the oxidation affected zone. Therefore, while configurational entropy stabilizes the solid solution phase in the base material, this effect is not expected to hold in the oxidation affected zone during high temperature oxidation.

## 6.0 Acknowledgements

This work is supported by the U.S. Office of Naval Research MURI program (grant no. N00014-15- 1-2863) and the Virginia Space Grant Consortium Graduate Research Fellowship. The authors would also like to thank Professors Bill Fahrenholtz (Missouri University of Science and Technology), Bill Soffa, Bi-Cheng Zhou (University of Virginia) and Christina Rost (James Madison University) for helpful discussions.